\documentclass[reqno,12pt,a4paper]{amsart}

\usepackage{mathrsfs}
\usepackage{amssymb}
\usepackage{amsfonts}
\usepackage{latexsym}
\usepackage{amsthm}

\usepackage{graphicx}
\def\lb{\label}

\newcommand{\er}[1]{\textrm{(\ref{#1})}}

\begin{document}


\renewcommand{\theequation}{\arabic{section}.\arabic{equation}}
\theoremstyle{plain}
\newtheorem{theorem}{\bf Theorem}[section]
\newtheorem{lemma}[theorem]{\bf Lemma}
\newtheorem{corollary}[theorem]{\bf Corollary}
\newtheorem{proposition}[theorem]{\bf Proposition}
\newtheorem{definition}[theorem]{\bf Definition}
\newtheorem{remark}[theorem]{\it Remark}

\def\a{\alpha}  \def\cA{{\mathcal A}}     \def\bA{{\bf A}}  \def\mA{{\mathscr A}}
\def\b{\beta}   \def\cB{{\mathcal B}}     \def\bB{{\bf B}}  \def\mB{{\mathscr B}}
\def\g{\gamma}  \def\cC{{\mathcal C}}     \def\bC{{\bf C}}  \def\mC{{\mathscr C}}
\def\G{\Gamma}  \def\cD{{\mathcal D}}     \def\bD{{\bf D}}  \def\mD{{\mathscr D}}
\def\d{\delta}  \def\cE{{\mathcal E}}     \def\bE{{\bf E}}  \def\mE{{\mathscr E}}
\def\D{\Delta}  \def\cF{{\mathcal F}}     \def\bF{{\bf F}}  \def\mF{{\mathscr F}}
\def\c{\chi}    \def\cG{{\mathcal G}}     \def\bG{{\bf G}}  \def\mG{{\mathscr G}}
\def\z{\zeta}   \def\cH{{\mathcal H}}     \def\bH{{\bf H}}  \def\mH{{\mathscr H}}
\def\e{\eta}    \def\cI{{\mathcal I}}     \def\bI{{\bf I}}  \def\mI{{\mathscr I}}
\def\p{\psi}    \def\cJ{{\mathcal J}}     \def\bJ{{\bf J}}  \def\mJ{{\mathscr J}}
\def\vT{\Theta} \def\cK{{\mathcal K}}     \def\bK{{\bf K}}  \def\mK{{\mathscr K}}
\def\k{\kappa}  \def\cL{{\mathcal L}}     \def\bL{{\bf L}}  \def\mL{{\mathscr L}}
\def\l{\lambda} \def\cM{{\mathcal M}}     \def\bM{{\bf M}}  \def\mM{{\mathscr M}}
\def\L{\Lambda} \def\cN{{\mathcal N}}     \def\bN{{\bf N}}  \def\mN{{\mathscr N}}
\def\m{\mu}     \def\cO{{\mathcal O}}     \def\bO{{\bf O}}  \def\mO{{\mathscr O}}
\def\n{\nu}     \def\cP{{\mathcal P}}     \def\bP{{\bf P}}  \def\mP{{\mathscr P}}
\def\r{\rho}    \def\cQ{{\mathcal Q}}     \def\bQ{{\bf Q}}  \def\mQ{{\mathscr Q}}
\def\s{\sigma}  \def\cR{{\mathcal R}}     \def\bR{{\bf R}}  \def\mR{{\mathscr R}}
\def\S{\Sigma}  \def\cS{{\mathcal S}}     \def\bS{{\bf S}}  \def\mS{{\mathscr S}}
\def\t{\tau}    \def\cT{{\mathcal T}}     \def\bT{{\bf T}}  \def\mT{{\mathscr T}}
\def\f{\phi}    \def\cU{{\mathcal U}}     \def\bU{{\bf U}}  \def\mU{{\mathscr U}}
\def\F{\Phi}    \def\cV{{\mathcal V}}     \def\bV{{\bf V}}  \def\mV{{\mathscr V}}
\def\P{\Psi}    \def\cW{{\mathcal W}}     \def\bW{{\bf W}}  \def\mW{{\mathscr W}}
\def\o{\omega}  \def\cX{{\mathcal X}}     \def\bX{{\bf X}}  \def\mX{{\mathscr X}}
\def\x{\xi}     \def\cY{{\mathcal Y}}     \def\bY{{\bf Y}}  \def\mY{{\mathscr Y}}
\def\X{\Xi}     \def\cZ{{\mathcal Z}}     \def\bZ{{\bf Z}}  \def\mZ{{\mathscr Z}}
\def\O{\Omega}

\newcommand{\mc}{\mathscr {c}}

\newcommand{\gA}{\mathfrak{A}}          \newcommand{\ga}{\mathfrak{a}}
\newcommand{\gB}{\mathfrak{B}}          \newcommand{\gb}{\mathfrak{b}}
\newcommand{\gC}{\mathfrak{C}}          \newcommand{\gc}{\mathfrak{c}}
\newcommand{\gD}{\mathfrak{D}}          \newcommand{\gd}{\mathfrak{d}}
\newcommand{\gE}{\mathfrak{E}}
\newcommand{\gF}{\mathfrak{F}}           \newcommand{\gf}{\mathfrak{f}}
\newcommand{\gG}{\mathfrak{G}}           
\newcommand{\gH}{\mathfrak{H}}           \newcommand{\gh}{\mathfrak{h}}
\newcommand{\gI}{\mathfrak{I}}           \newcommand{\gi}{\mathfrak{i}}
\newcommand{\gJ}{\mathfrak{J}}           \newcommand{\gj}{\mathfrak{j}}
\newcommand{\gK}{\mathfrak{K}}            \newcommand{\gk}{\mathfrak{k}}
\newcommand{\gL}{\mathfrak{L}}            \newcommand{\gl}{\mathfrak{l}}
\newcommand{\gM}{\mathfrak{M}}            \newcommand{\gm}{\mathfrak{m}}
\newcommand{\gN}{\mathfrak{N}}            \newcommand{\gn}{\mathfrak{n}}
\newcommand{\gO}{\mathfrak{O}}
\newcommand{\gP}{\mathfrak{P}}             \newcommand{\gp}{\mathfrak{p}}
\newcommand{\gQ}{\mathfrak{Q}}             \newcommand{\gq}{\mathfrak{q}}
\newcommand{\gR}{\mathfrak{R}}             \newcommand{\gr}{\mathfrak{r}}
\newcommand{\gS}{\mathfrak{S}}              \newcommand{\gs}{\mathfrak{s}}
\newcommand{\gT}{\mathfrak{T}}             \newcommand{\gt}{\mathfrak{t}}
\newcommand{\gU}{\mathfrak{U}}             \newcommand{\gu}{\mathfrak{u}}
\newcommand{\gV}{\mathfrak{V}}             \newcommand{\gv}{\mathfrak{v}}
\newcommand{\gW}{\mathfrak{W}}             \newcommand{\gw}{\mathfrak{w}}
\newcommand{\gX}{\mathfrak{X}}               \newcommand{\gx}{\mathfrak{x}}
\newcommand{\gY}{\mathfrak{Y}}              \newcommand{\gy}{\mathfrak{y}}
\newcommand{\gZ}{\mathfrak{Z}}             \newcommand{\gz}{\mathfrak{z}}

\def\ve{\varepsilon}   \def\vt{\vartheta}    \def\vp{\varphi}    \def\vk{\varkappa}

\def\A{{\mathbb A}} \def\B{{\mathbb B}} \def\C{{\mathbb C}}
\def\dD{{\mathbb D}} \def\E{{\mathbb E}} \def\dF{{\mathbb F}} \def\dG{{\mathbb G}} \def\H{{\mathbb H}}\def\I{{\mathbb I}} \def\J{{\mathbb J}} \def\K{{\mathbb K}} \def\dL{{\mathbb L}}\def\M{{\mathbb M}} \def\N{{\mathbb N}} \def\dO{{\mathbb O}} \def\dP{{\mathbb P}} \def\R{{\mathbb R}}\def\S{{\mathbb S}} \def\T{{\mathbb T}} \def\U{{\mathbb U}} \def\V{{\mathbb V}}\def\W{{\mathbb W}} \def\X{{\mathbb X}} \def\Y{{\mathbb Y}} \def\Z{{\mathbb Z}}


\def\la{\leftarrow}              \def\ra{\rightarrow}            \def\Ra{\Rightarrow}
\def\ua{\uparrow}                \def\da{\downarrow}
\def\lra{\leftrightarrow}        \def\Lra{\Leftrightarrow}


\def\lt{\biggl}                  \def\rt{\biggr}
\def\ol{\overline}               \def\wt{\widetilde}
\def\no{\noindent}


\let\ge\geqslant                 \let\le\leqslant
\def\lan{\langle}                \def\ran{\rangle}
\def\/{\over}                    \def\iy{\infty}
\def\sm{\setminus}               \def\es{\emptyset}
\def\ss{\subset}                 \def\ts{\times}
\def\pa{\partial}                \def\os{\oplus}
\def\om{\ominus}                 \def\ev{\equiv}
\def\iint{\int\!\!\!\int}        \def\iintt{\mathop{\int\!\!\int\!\!\dots\!\!\int}\limits}
\def\el2{\ell^{\,2}}             \def\1{1\!\!1}
\def\wh{\widehat}
\def\bs{\backslash}
\def\intl{\int\limits}

\def\na{\mathop{\mathrm{\nabla}}\nolimits}
\def\sh{\mathop{\mathrm{sh}}\nolimits}
\def\ch{\mathop{\mathrm{ch}}\nolimits}
\def\where{\mathop{\mathrm{where}}\nolimits}
\def\all{\mathop{\mathrm{all}}\nolimits}
\def\as{\mathop{\mathrm{as}}\nolimits}
\def\Area{\mathop{\mathrm{Area}}\nolimits}
\def\arg{\mathop{\mathrm{arg}}\nolimits}
\def\const{\mathop{\mathrm{const}}\nolimits}
\def\det{\mathop{\mathrm{det}}\nolimits}
\def\diag{\mathop{\mathrm{diag}}\nolimits}
\def\diam{\mathop{\mathrm{diam}}\nolimits}
\def\dim{\mathop{\mathrm{dim}}\nolimits}
\def\dist{\mathop{\mathrm{dist}}\nolimits}
\def\Im{\mathop{\mathrm{Im}}\nolimits}
\def\Iso{\mathop{\mathrm{Iso}}\nolimits}
\def\Ker{\mathop{\mathrm{Ker}}\nolimits}
\def\Lip{\mathop{\mathrm{Lip}}\nolimits}
\def\rank{\mathop{\mathrm{rank}}\limits}
\def\Ran{\mathop{\mathrm{Ran}}\nolimits}
\def\Re{\mathop{\mathrm{Re}}\nolimits}
\def\Res{\mathop{\mathrm{Res}}\nolimits}
\def\res{\mathop{\mathrm{res}}\limits}
\def\sign{\mathop{\mathrm{sign}}\nolimits}
\def\span{\mathop{\mathrm{span}}\nolimits}
\def\supp{\mathop{\mathrm{supp}}\nolimits}
\def\Tr{\mathop{\mathrm{Tr}}\nolimits}
\def\BBox{\hspace{1mm}\vrule height6pt width5.5pt depth0pt \hspace{6pt}}


\newcommand\nh[2]{\widehat{#1}\vphantom{#1}^{(#2)}}
\def\dia{\diamond}

\def\Oplus{\bigoplus\nolimits}



\def\qqq{\qquad}
\def\qq{\quad}
\let\ge\geqslant
\let\le\leqslant
\let\geq\geqslant
\let\leq\leqslant
\newcommand{\ca}{\begin{cases}}
\newcommand{\ac}{\end{cases}}
\newcommand{\ma}{\begin{pmatrix}}
\newcommand{\am}{\end{pmatrix}}
\renewcommand{\[}{\begin{equation}}
\renewcommand{\]}{\end{equation}}
\def\eq{\begin{equation}}
\def\qe{\end{equation}}
\def\[{\begin{equation}}
\def\bu{\bullet}

\title[{Resonances for 1D massless Dirac operators}]
{Resonances for 1D massless Dirac operators}

\date{\today}

\author[ Alexei Iantchenko]{ Alexei Iantchenko}
\address{Malm{\"o} h{\"o}gskola, Teknik och samh{\"a}lle, 205 06
  Malm{\"o}, Sweden, email: ai@mah.se }
\author[Evgeny Korotyaev]{Evgeny Korotyaev}
\address{Mathematical Physics Department, Faculty of Physics, Ulianovskaya 2,
St. Petersburg State University, St. Petersburg, 198904,
 and Pushkin Leningrad State University, Russia,
 \ korotyaev@gmail.com,
}


\begin{abstract}
\no We consider the 1D massless Dirac operator    on the real line with compactly supported potentials.
We study  resonances as the poles of scattering matrix or equivalently as the zeros of modified Fredholm determinant.
We obtain the following properties of the resonances:
1) asymptotics of counting function,
2)  estimates on the resonances and the forbidden domain,
3)  the trace formula in terms of resonances.
 \\ \\
\end{abstract}

\maketitle


\vskip 0.25cm
\section {Introduction and main results}
\setcounter{equation}{0}

We consider the 1D massless Dirac operator $H$  acting in the Hilbert space $L^2(\R )\os L^2(\R )$ and given by
$$
H=-iJ {d\/ dx}+ V(x),\ \ \ \  J =\ma 1&0\\ 0&-1\am, \ \ \ \ \
V= \ma 0&q\\ \ol q & 0\am.
$$
 Here  $q\in L^1(\R)\cap L^2(\R)$ is a complex-valued function. In order to define resonances we will need to suppose that $q$ has compact support and satisfy the following hypothesis:

\no {\bf Condition A}. {\it A complex-valued function $q\in L^2(\R)$
and $\supp q\ss [0,\g],$ for some $\g>0$,
 where  $[0,\g]$ is the convex
hull of the support of $q$.}

It is well known (see \cite{DEGM}, \cite{ZMNP})   that the operator $H$ is
self-adjoint and its spectrum  is purely absolutely
continuous and is given by the set $\R$.

We consider the Dirac equation for a vector
valued function $f(x)$
\[
\lb{1}
-i Jf'+V  f=\l f, \ \ \ \l \in \C,\ \ \ \
f(x)= f_1(x)e_++f_2(x)e_-,  \ \ \ e_+= \ma 1\\0\am,
   \   e_-= \ma 0\\1\am,
\]
where $f_1, f_2$ are  the functions in $x\in\R $. System \er{1} is
also known as the Zakhorov-Shabat system (see \cite{DEGM},
\cite{ZMNP}). Define the fundamental solutions $\p_{\pm}, \vp_{\pm},
$ of (1.1) under the following conditions
$$
\p^{\pm}(x,\l )=e^{\pm i\l x}e_{\pm},\ \ \ \  x>\g;\ \ \  \ \  \ \ \
\ \ \ \vp^{\pm}(x,\l )=e^{\pm i\l x}e_{\pm},\ \ \ \  x<0.
$$
  Define the  functions
\[\lb{ab}
a(\l)=\det (\p^+ (x,\l ),\vp^-(x,\l )) , \ \ \ \
b(\l)=\det (\vp^-(x,\l ),\p^-(x,\l )),
\]
where $\det (f,g)$ is the Wronskian for two vector-valued functions
$f, g$.

 {\bf Below we consider all functions and the resolvent in
upper-half plane $\C_+$ and we will obtain their analytic
continuation into the whole complex plane $\C$.} Note that we can
consider all functions and the resolvent in lower-half plane $\C_-$
and to obtain their analytic continuation into the whole complex
plane $\C$. The Riemann surface of the resolvent for the Dirac
operator consists of two disconnected sheets $\C$. In the case of
the Schr\"odinger operator
 the corresponding Riemann surface is the Riemann surface  of the function
 $\sqrt \l$.

 {\bf The zeros of $a(\l)$ in $\C_-$ are called {\em resonances}
 with multiplicities as zeros of function $a(\l).$}

Before to proceed with our results, we need to give a short introduction to the subject.

Resonances, from a physicists
point of view, were first studied by Regge in 1958 (see \cite{R58}). Since then,
 the properties of  resonances has been the object of intense study and we refer to   \cite{SZ91} for the mathematical approach in the multi-dimensional case  and references
given there. In the multi-dimensional Dirac case resonances were studied locally in \cite{HB92}.
We discuss the global properties of resonances in the one-dimensional case. A lot of papers are
devoted to the resonances for the 1D Schr\"odinger operator, see
Froese \cite{F97}, Korotyaev \cite{K04}, Simon \cite{S00}, Zworski
\cite{Z87} and references given there. We recall that Zworski \cite{Z87}
obtained the first results about the asymptotic distribution of
resonances for the Schr\"odinger operator with compactly supported
potentials on the real line. Different properties of resonances were
determined in \cite{H99},  \cite{K11}, \cite{S00} and \cite{Z87}. Inverse problems (characterization,
recovering, plus uniqueness) in terms of resonances were solved by
Korotyaev  for the Schr\"odinger operator with a compactly supported
potential on the real line \cite{K05} and the half-line \cite{K04}.

The "local resonance" stability problems were considered in
\cite{K04s}, \cite{MSW10}.

However, we know only one paper \cite{K12} about the
resonances for the Dirac operator $H$ on the real line.
In particular, for each $p>1$
 the estimates of resonances in terms of potentials are obtained:
$$ \sum_{\Im \l_n< 0} {1\/|\l_n-i|^{p}}\le   {C Y_p\/\log
2}\rt({4\g\/\pi}+\int_\R |q(x)|dx\rt),
$$
where $C\le 2^5$ is an absolute  constant and  $Y_p=\sqrt \pi{
\G({p-1\/2})\/\G({p\/2})},$   and $\G$ is the Gamma function.

Inverse scattering theory for the Zakharov-Shabat systems were
developed for the investigation of NLS, see \cite{FT87}, \cite{DEGM},
\cite{ZMNP}. In \cite{Gr92} Grebert studies the inverse scattering problem for the  Dirac operator on the real line. In \cite{IK2} we give the properties of bound states and resonances for the Dirac operator with mass $m>0$ on the half-line. In \cite{IK3} we describe the properties of graphene with localized impurities modeled by the two-dimensional Dirac operator with compactly supported radial potential.
We address the inverse resonance problem for 1D Dirac operators in \cite{IK4}.

In this paper we study resonances for the massless Dirac operator. This analysis is based on the properties of functions $a,$ $b$ defined in (\ref{ab}).
We will show that functions $a, b$ are entire and
\[
\lb{asa}
a(i\e)=1+ o(1)\qq \mbox{as}  \qq \e\to\iy .
\]
 All zeros of $a(\l)$
  lie in $\C_-.$  We denote by $(\l_n)_1^{\iy}$
the sequence of zeros in $\C_-$ of $a$ (multiplicities counted by repetition),
so arranged that
$$
0<|\l_1|\leq |\l_2|\leq |\l_2|\leq \dots
$$
and let
$\l_n=\m_n+i\e_n, n\geq 1$. The massless  Dirac operator with $q\equiv 0$ we denote by $H_0$.
The scattering matrix $\cS$ for the pair $H, H_0$ has the following form
$$
\cS(\l )={1\/a(\l)}\ma 1& -\ol b(\l)\\ b(\l)& 1\am,
\qqq \l\in \R.
$$
Here $1/a$ is the transmission coefficient and $-\ol b/a$ (or  $b/a$) is
the right (left) reflection coefficient.
Due to \er{asa} we take the unique branch $\log a(\l)=o(1)$ as $\l=i\e,$ $\e \to \iy$.

We define the function
$$
\log a(\l,q)=\n (\l,q)+i\f_{\rm sc}(\l,q),\ \ \ \ \f_{\rm sc} (\l,q)=\arg a(\l ,q),
\ \ \ \ \n (\l,q)=\log |a(\l ,q)|, \ \ \ \ \l\in \C_+ ,
$$
where the function $\f$ is called the scattering phase (or
the spectral shift function, see [Kr]) and the function $\n$
is called the action variable for  the non-linear Schro\"odinger equation on the real line (see \cite{ZMNP}). The scattering matrix $S(\l)$ is unitary and  we have the identities
$$
|a(\l)|^2- |b(\l)|^2=1, \ \ \ \l \in \R ,
$$
$$
\det \cS(\l )=e^{-i2\arg a(\l)}=e^{-i2 \f_{\rm sc} (\l )},\qq\l\in\R.
$$
If $q''\in L^2(\R ) $ then we have the following asymptotic estimate (see \cite{ZMNP})
\[
\lb{asa2}
 i\log a(\l)=-{Q_0\/\l }-{Q_1\/\l ^{2}}-{Q_2+o(1)\/\l ^{3}},
\ \ \ \ \l =i\e ,\ \ \e\to \iy ,
\]
where $Q_j={1\/\pi}\int _{\R}\l^j\log |a(\l)|\,d\l, j=0,1,\ldots ,$
$$
Q_j=2^{-j}\cH_j,\qqq j=0,1,2,
$$
$$
\begin{aligned}
\cH_0={1\/2}\int_\R |q(x)|^2dx,\qqq
\cH_1={1\/2}\int_\R q'(x) \overline{q}(x)dx,
\qq
\cH_2={1\/2}\int_\R(|q'(x)|^2+|q(x)|^4)dx.
\end{aligned}
$$
Here $\cH_j$ are hierarchy of the defocussing cubic non-linear Schr\"odinger equation (dNLS) on the real line given by
$$
-i{\pa\p \/\pa t}=-\p _{xx}+2|\p |^2\p.
$$

 The main goal of our paper is to describe the properties of the resonances and to determine
the trace formula in terms of resonances.  We achieve this goal by studying  the
properties of function $a(\l)$ which is entire function of exponential type with zeros in
$\C_-.$

We introduce the modified Fredholm determinant (see \cite{GK69})
 as follows. Using the factorization of potential $V,$ we
introduce the operator valued function (the sandwich operator)
$Y_0(\l)$ by
$$
Y_0(\l)=V_2R_0(\l)V_1,   \qq \mbox{where} \qq
V=V_1V_2,\,\, V_2=|q|^{1\/2}I_2.
$$
 Observing  that $Y_0(\l)$ is in the Hilbert-Schmidt class $\cB_2$ but
 not in the trace class $\cB_1$ (explained in the beginning of Section \ref{s-ModFrDet}),
 we define the modified Fredholm determinant $D(\l)$ by
$$
D(\l)=\det\left[ (I+Y_0(\l)) e^{-Y_0(\l)}\right], \qqq \l\in \C_+.
$$
Then the function $D(\cdot)$ is well-defined in $\C_+.$

We define the space $L^p(\R),$ $p\ge 1,$ equipped  with the standard norm
$
\|f\|_p=\left(\int_\R|f(x)|^pdx\right)^{\frac{1}{p}}.
$

Let $\cH_+^2$ denote the Hardy class of functions $g$ which are analytic in $\C_+$ and satisfy $$\sup_{y>0}\int_\R|g(x+iy)|^2dx <\infty.$$

  We formulate now the main result about
the function $a(\l).$

\begin{theorem}\lb{T1}
Let $q\in L^1(\R)\cap L^2(\R)$. Then function $a$ and the
determinant $D$ are analytic in $\C_+,$  continuous up to the
real line  and satisfy
\[
\lb{a=D} a= D,
\]
\[\lb{Hardy}
\log D(\cdot)=\log a(\cdot)\in \cH_+^2.\]

Moreover, if in addition $q'\in L^1(\R)$ then
\[
\lb{2}
 i\log a(\l)=-{\|q\|_2^2+o(1)\/2\l }
\qqq \as \qq \Im \l\to \iy .
\]
\end{theorem}
{\bf Remark.} 1) To the best of our knowledge, the important identity (\ref{a=D}) is new in the settings of Dirac systems. We will stress on the fact that for the massless Dirac operator there is no factor of proportionality in the identity. In the massive case (see \cite{IK2, IK3}) the situation is different.

2) The proof of Theorem \ref{T1} follows from Lemma \ref{TD1} and is given in Section \ref{s-ModFrDet}.

We determine the asymptotics of the counting function. We denote the
number of zeros of a function $f$ having modulus  $\leq r$ by $\cN
(r,f)$, each zero being counted according to its multiplicity.

\begin{theorem}
\lb{T2}  Assume that potential $q$ satisfies Condition A. Then $a(\cdot)$
has an analytic continuation from $\C_+$ into the whole complex
plane $\C$ and satisfies:
\[
\lb{counting}
\cN(r,a)= {2r\g\/ \pi }(1+o(1))\qqq as \qqq r\to\iy.
\]
Moreover, for each $\d >0$ the number of zeros of $a$ with modulus $\leq r$
lying outside both of the two sectors $|\arg z |<\d ,$ $|\arg z -\pi
|<\d$ is $o(r)$ for large $r$.

\end{theorem}
{\bf Remark.} 1) Zworski  obtained in \cite{Z87} similar results for the Schr{\"o}dinger operator with compactly supported potentials on the real line.

2) Our proof follows from Proposition \ref{Prop2} and Levinson Theorem \ref{th-Levinson}.

The analytic properties of function $a$ imply  estimates of resonances in terms of the potential.
\begin{theorem}
\lb{T3} Assume that potential $q$ satisfies Condition A and   $q'\in
L^1(\R).$  Let $\l_n\in\C_-,$ $n\geq 1,$ be any zero of $a(\l)$ in $\C_-$ (i.e. resonance). Then
\[\lb{log_neigh}
\left|\l_n^2 +\frac{i}{2}\l_n\|q\|_2^2\right|\leq
C_1e^{-2\g\Im\l_n},
\]
where the  constant
\[\lb{forbdo}
C_1=\sup_{\l\in\R}\left|
\l^2\left(a(\l)-1+\frac{1}{2i\l}\|q\|_2^2\right)\right| <\infty.
\]
In particular, for any $A>0,$ there are only finitely many resonances in the region
$$\{\Im\l \geq -A-\frac{1}{\gamma}\log|\Re\l_n|\}.$$
\end{theorem}
{\bf Remark.} 1) The similar results for the Schr{\"o}dinger operator were obtained in \cite{K04}.

2) Estimate (\ref{forbdo}) describes the forbidden domain for the resonances.

3) The proof of the theorem follows from Corollary \ref{Cor3.3.} using Proposition \ref{Prop2}, Lemma \ref{l-smooth q} and asymptotics (\ref{uniform_bound}).

We determine the trace formulas in terms of resonances   for the Dirac operator.

\begin{theorem}
\lb{T4}  Assume that potential $q$ satisfies Condition A. Let
$f(\l)=\hat{\vp}(\l),$ for $\vp\in C_0^\infty,$ and let
$\l_n$ be a resonance and $\f_{\rm sc}'(\l)$ the scattering phase,
then
\[\lb{TrF1}
\Tr (f(H)-f(H_0))=-\int_\R f(\l)\f_{\rm sc}'(\l)d\l=\sum_{\n\geq 1} f(\l_n), \]
\[\lb{sc_phase}
\f_{\rm sc}'(\l)=\frac{1}{\pi}\sum_{n\ge1}\frac{\Im\l_n}{|\l-\l_n|^2}, \qq \l\in\R.\]
\[\lb{TrF3}
\Tr (R(\l)-R_0(\l))=-i\g -\lim_{r\to +\infty} \sum_{|\l_n|\le
r}\frac{1}{\l-\l_n},
\]
where the series converge uniformly in every bounded subset on the
plane by condition (\ref{sumcond}).
\end{theorem}
{\bf Remark.} These results are  similar to the 1D Schr{\"o}dinger  case (see Korotyaev \cite{K04}, \cite{K05}) and in 3D (see \cite{IK11}).

The plan of our paper is as follows.  In Section \ref{s-Cart} we recall some
results about entire functions and  prove Theorem
\ref{T3} referring to the results obtained in Section \ref{s-Dodd improved}. In Section \ref{s-Dirac systems}
 we describe the properties of fundamental solutions and  prove Theorem
\ref{T2}. In Section \ref{s-Dodd improved} we obtain uniform estimates on the Jost solution as $\l\rightarrow\infty$  under the condition that $q'\in L^1(\R).$  In Section \ref{s-determinant} we prove useful Hilbert-Schmidt estimates for the "sandwiched" free resolvent . In Section \ref{s-ModFrDet} we give the properties of the modified Fredholm determinant and prove Theorems \ref{T1} and \ref{T4}.

\section{ Cartwright class of entire functions}\lb{s-Cart}
\setcounter{equation}{0}

 In this section we will prove  Theorem \ref{T3}. The proof is based on some well-known facts from the theory of entire functions which we recall here. We  mostly follow \cite{Koo81}. We denote the
number of zeros of function $f$ having modulus  $\leq r$ by $\cN
(r)$, each zero being counted according to its multiplicity. We
sometimes write $\cN (r,f)$ instead of $\cN (r)$ when several
functions are being dealt with. An entire function $f(z)$ is said to
be of exponential type if there is a constant $A$ such that
$|f(z)|\leq\const e^{A|z|}$ everywhere.  The infimum of the set of
$A$ for which such inequality   holds is called the type of
$f$. For each exponential type function $f$ we define the types $\r
_{\pm}(f)$ in $\C_\pm $ by
$$
\r _{\pm}(f)\equiv \lim \sup_{y\to \iy} {\log |f(\pm iy)|\/y} .
$$
Fix $\rho >0.$ We introduce the class of exponential type  functions

\no {\bf Definition.} {\it Let $\cE_\delta (\rho),$ $\d >0,$ denote
the space of exponential type  functions $f$, which satisfy the
following conditions:

\no i)  $\r_+(f)=0$ and $ \r_-(f)=\rho$,

\no ii) $f(z)$ does not have  zeros in $\C_+$,

\no iii) $f\in  L^\infty (\R),$

\no iv) $|f(x)|\geq \delta$ for all $x\in \R $. }

 The function $f$ is
said to  belong    to the Cartwright class  if $f$ is entire, of
exponential type, and the following conditions hold true:
$$
\int_\R\frac{\log(1+|f(x)|)}{1+x^2}dx
<\infty,\qq\rho_+(f)=0,\qq\rho_-(f)=\rho>0,
$$   for some $\rho>0.$

Assume $f$ belong to the Cartwright class and denote by
$(z_n)_{n=1}^\infty$ the sequence of its zeros $\neq 0$ (counted
with multiplicity), so arranged that $0<|z_1|\leq|z_2|\leq\ldots.$
Then we have the Hadamard factorization
\begin{equation}\lb{Hadfact}
f(z)=Cz^m e^{i\rho z/2 }\lim_{r\rightarrow\infty}\prod_{|z_n|\leq
r}\left(1-\frac{z}{z_n}\right),\qq C=\frac{f^{(m)}(0)}{m!} ,
\end{equation} for some integer $m,$ where
the product converges uniformly in every bounded disc and
\begin{equation}
\lb{sumcond}
\sum {|\Im z_n |\/|z_n|^2} <\infty.
\end{equation}

Given an entire function $f$, let us denote by $\cN_{+}(r,f)$
the number of its zeros with real part $\geq 0$ having modulus $\leq r$,
and by $\cN_{-}(r,f)$ the number of its zeros with real part $< 0$
having modulus $\leq r$.
As usual,  $\cN (r,f)=\cN_{-}(r,f)+\cN_{+}(r,f)$ is the total number of  zeros
of $f$  with modulus $\leq r$, and multiple zeros  $f$ are counted
accoding to their multiplicities in reckoning the quantities
$\cN_{-}(r,f), \cN_{+}(r,f)$ and $\cN (r,f)$. We need the following
 well known result (see  \cite{Koo81}, page 69).

\begin{theorem}[Levinson]\lb{th-Levinson} Let the function $f$ belong to the Cartwright class for some $\rho >0.$
Then
\[
  \cN_{\pm }(r,f)={\rho\, r\/2 \pi }(1+o(1))\qq\mbox{as}\qq r\to \iy .
\]
For each $\d >0$ the number of zeros of $f$ with modulus $\leq r$
lying outside both of the two sectors $|\arg z | , |\arg z -\pi |<\d$
is $o(r)$ for large $r$.
\end{theorem}

Below we will use some arguments from the paper \cite{K04}, where some
properties of resonances were proved for the Schr\"odinger operators. In order to adapt the formulas to our settings we write $\rho=2\gamma,$ $\gamma >0.$ In order to prove Theorem \ref{T2} we need

\begin{lemma}
\lb{L3.2} Let $f\in \cE_{\d}(2\g)$ for some $\delta\in [0,1]$ and $\gamma >0.$ Assume that for some $p\geq 0$
there exists a polynomial $G_p(z)=1+\sum _1^pd_nz^{-n}$ and a
constant $C_p$ such that
\[
\lb{3.10} C_p=\sup_{x\in \R }  |x^{p+1}(f(x)-G_p(x))|<\iy.
\]
Then for each zero $z_n, n\geq 1,$ the following estimate holds
true:
\[
\lb{3.11}
  |G_p(z_n))|\leq C_p|z_n|^{-p-1}e^{-2\g y_n},\qq y_n=\Im z_n.
\]
\end{lemma}
{\bf Proof.} We take the function
$f_p(z)=z^{p+1}(f(z)-G_p(z))e^{-i2\g z}$. By condition, the function
$f_p$ satisfies the estimates

\no 1) $|f_p(x)|\leq C_p$ for $x\in \R $,

\no 2) $\log |f_p(z)|\leq {\mathcal O}(|z|)$ for large $z\in \C_- $

\no 3) $\lim\sup_{y\to\iy }y^{-1} \log |f_p(-iy)|=0$

Then the Phragmen-Lindel{\"o}f Theorem (see \cite{Koo81},   page 23) implies $|f_p(z)|\leq C_p$ for $z\in \C_- $. Hence at $z=z_n$ we obtain
\[
\lb{3.13} |z^{p+1}G_p(z)e^{-i2\g
z}|=|f_p(z)|=|z^{p+1}(f(z)-G_p(z))e^{-i2\g z}|\leq C_p,
\]
which yields \er{3.11}. \hfill\BBox

\no    \begin{corollary}\lb{Cor3.3.}
 Let $f\in \cE_{\d}(2\g)$ for some $\delta\in [0,1]$ and $\gamma >0.$ and let $z_n, n\ge 1,$ be zeros of $f$.\\
 i) Assume that $C_0=\sup_{x\in \R} |x(f(x)-1)|<\iy$.
 Then each zero $z_n, n\geq 1,$ satisfies
\[
\lb{3.14a}
  |z_n|\leq C_0e^{-2\g y_n}.
\]
ii) Assume that $C_1=\sup_{x\in \R} |x^2f(x)-x^2-Ax|<\iy$ for
some $A$. Then each zero $z_n, n\geq 1,$ satisfies
\[
\lb{3.15}
  |z_n(z_n+A)|\leq C_1e^{-2\g y_n}.
\]
\end{corollary}
{\bf Proof of Theorem \ref{T3}.}
Note that in Proposition \ref{Prop2} it is proved that the inverse of the transmission coefficient $a(\l)$ belongs to $\cE_{1}(2\g).$ Moreover, if $q$ satisfies Condition A and $q'\in L^1(\R),$ then $a(\l)$ satisfy  uniform bound (\ref{uniform_bound}), Lemma \ref{l-smooth q}, and therefore the conditions of Corollary \ref{Cor3.3.} are satisfied with $A=\frac{i}{2}\|q\|_2^2.$ \hfill\BBox

\section {Dirac systems }\lb{s-Dirac systems}
\setcounter{equation}{0}

\subsection{Preliminaries }
We consider the Dirac system  (\ref{1}) for a vector valued
function $f(x)= f_1(x)e_++f_2(x)e_-,$ where $f_1, f_2$ are  the functions of $x\in\R :$
\[\lb{1.2}
 \left\{\begin{array}{c}
           -if_1'+qf_2=\l f_1 \\
           if_2'+\overline{q} f_1=\l f_2
         \end{array}\right. \ \ \ \l \in \C.
\]

Here $q\in L^1(\R)\cap L^2(\R)$ is a complex-valued function.

Note that if the function  $f(x,\l)=(f_1(x,\l),f_2(x,\l))^{\rm T}$
is solution of (\ref{1.2}) with $\l\in\C,$ then
$\wt{f}(x,\l):=(\overline{f}_2(x,\overline{\l}),\overline{f}_1(x,\overline{\l}))^{\rm
T}$ is also the solution of (\ref{1.2}) with the same $\l.$

Define the fundamental solutions $\p_{\pm}, \vp_{\pm}, $ of (\ref{1.2}) satisfying
the following conditions
$$
\p^{\pm}(x,\l )=e^{\pm i\l x}e_{\pm},\ \ \ \  x>\g;\ \ \  \ \  \ \ \
\ \ \ \vp^{\pm}(x,\l )=e^{\pm i\l x}e_{\pm},\ \ \ \  x < 0.
$$ Then
$$
\det (\p^+ (x,\l ),\p^-(x,\l ))=\det (\vp^+ (x,\l ),\vp^-(x,\l ))=1,
$$$$
 \wt{\psi}^+(x,\l)=\psi^- (x,\l),\qq\wt{\vp}^+(x,\l)=\vp^- (x,\l).
 $$

For $\l\in\R,$ we have
$$
  \vp^-(x,\l )=b(\l )\p^+(x,\l )+a(\l )\p^-(x,\l ),\qq \p^+(x,\l )=\wt{b}(\l )\vp^-(x,\l )+a(\l )\vp^+(x,\l ),
$$
where
$$
a(\l)=\det (\p^+ (x,\l ),\vp^-(x,\l )),\qq  b(\l)=\det (\vp^-(x,\l ),\p^-(x,\l ))$$ and
$$\qq \wt{b}(\l)=\det (\vp^+(x,\l),\p^+(x,\l))=\det (\wt{\vp}^-,\wt{\p}^-)=-\overline{b}(\overline{\l}).$$

Using the property that if $f$ is solution of (\ref{1.2}) with $(\l,q)$ then $\overline{f}$ is solution of (\ref{1.2}) with $(-\overline{\l},-\overline{q})$ we get that
\begin{equation}\lb{conjugation}
\overline{a(\l,q)}=a(-\overline{\l},-\overline{q}),\qq \overline{b(\l,q)}=b(-\overline{\l},-\overline{q}).
\end{equation}

We denote the operator with $q\equiv 0$  by $H_0$.
The scattering matrix $\cS$ for the pair $H, H_0$ has the following form
$$
\cS(\l )={1\/a(\l)}\ma 1& -\ol b(\l)\\ b(\l)& 1\am, \qqq \ \ \ R_-=
{\wt{b}\/ a}, \ \ \ R_+={b\/ a},
$$
here ${1\/ a}$ is the transmission coefficient and $R_\pm$ is the
right (left) reflection coefficient. Note that if $f=(f_1,f_2)^{\rm
T}$ is solution of (\ref{1.2}) with $\l\in\R$, then
$\wt{f}=(\overline{f}_2,\overline{f}_1)^{\rm T}$ is also the
solution of (\ref{1.2}) with the same $\l\in\R$. The S-matrix is
unitary, which implies the identity
\begin{equation}
\lb{identity-on-real}
|\det\cS(\l)|=|a(\l)|^2- |b(\l)|^2=1, \ \ \ \forall \l \in \R .
\end{equation}

\subsection{Properties of the fundamental solutions}
Now, we consider some properties  of the fundamental solutions
$\p^{\pm},  \vp^{\pm}$ of the Dirac system (\ref{1.2}) and
functions $a,\wt{b}$ for $\l\in\C.$ If function $q$ satisfies
Condition A, then
\[\lb{5.11}
a(\l )=\det (\p^+,\vp^-)=\p_1^{+}(0,\l ),\ \ \ \ b(\l )=\det
(\vp^-,\p^-)=-\p_1^{-}(0,\l )\] and \[\lb{btilda} \wt{b}(\l )=\det
(\vp^+,\p^+)=\p_2^{+}(0,\l ) .\]

The solutions $\p^{\pm}, \vp^{\pm}$ satisfy the following integral
equations:
\[
\lb{p} \p^{\pm}(x,\l )=e^{{\pm}i\l x}e_{\pm}+\int _x^\infty iJe^{i\l
(x-t)J} V(t) \p^{\pm}(t,\l )dt,
\]
\[
\lb{vp} \vp^{\pm}(x,\l )=e^{{\pm}i\l x}e_{\pm}-\int _0^xiJe^{i\l
(x-t)J} V(t)\vp^{\pm}(t,\l )dt,
\]
where
\begin{equation}\lb{5.6}
 iJe^{i\l (x-t)J} V(t)
= i\ma 0& q(t) e^{i\l (x-t)}\\ -\overline{q}(t)e^{-i\l (x-t)}&0\am .
\end{equation}
Using (\ref{5.6}) we obtain
$$
 \p_{1}^+(x,\l )=e^{i\l x}+i\int _x^\infty e^{i\l (x-t)}q(t) \p_{2}^+(t,\l )dt,
$$
$$
 \p_{2}^+(x,\l )=-i\int _x^\infty e^{-i\l (x-t)}\overline{q}(t) \p_{1}^+(t,\l )dt.
$$
Then
$$
 \p_{1}^+(x,\l )=e^{i\l x} + \int _x^\infty e^{i\l (x-t)}q(t)
\int _t^\infty e^{-i\l (t-s)}\overline{q}(s) \p_{1}^+(s,\l )dsdt,
$$
and we have the following equation for $\c=\p_{1}^+(x,\l )e^{-i\l x}$
\[
\lb{x0}
\begin{aligned}
&\c (x,\l )=1+\int_x^\infty q(t)\int_t^\infty e^{i2\l (s-t)}\ol{q}(s)\c
(s,\l )dsdt
=1+\int_x^\infty G(x,s,\l)\c (s,\l )ds,\\
&G(x,s,\l)=\ol{q}(s)\int_x^s e^{i2\l(s-t)}q(t)dt.
\end{aligned}
\]
Thus we have the power series in $q$
\[\label{series}
\c (x,\l )=1+\sum _{n\geq 1}\c_n(x,\l ),\ \ \ \ \c_n(x,\l
)=\int_x^\infty G(x,s,\l)\c_{n-1}(s,\l )ds,
\]
where $\c_0(\cdot ,\l )=1$.

\begin{lemma}\lb{chi_estimates} Suppose $q\in L^1(\R)\cap L^2(\R)$ and denote $\F(x)=\ch \int_x^\infty | q(s)|ds.$ Then the following facts hold true:

1) For each $x\in\R,$ the function $\c(x,\cdot)$ is continuous  in the closed half-plane $\Im\l\geq 0$ and entire in the open half-plane $\Im\l>0.$ For each $x\in\R$ and $\Im\l\geq 0,$ the functions $\c_n,$ $\c$ satisfy the following estimates:\\
\[
\lb{xn+}
  |\c_n(x,\l )|\leq \frac{1}{(2n)!}
  \left(\int_x^\infty|q(\t)|d\t\right)^{2n},\ \ \forall \ n\geq 1,\]
\[
\lb{x+}
  |\chi(x,\l)|\leq \F(x),
\]
\[\lb{x1+} |\chi(x,\l)-1|\leq \F(x)-1.\]

For all $\Im \l>0,$
\[
\lb{x2} |\chi(x,\l)-1|\le {\F(x)\/|\Im \l|^{1\/2}}\|q\|_2\|q\|_1.
\]

Moreover,
\[
\lb{L2norm} \int_\R|\chi(x,\l)-1|^2d\l\leq 4\pi
\left(\F(x)-1\right)\F(x)\|q\|_2^2.
\]

2) If $q$ satisfies Condition A, then for each $x\in\R,$ the function $\c(x,\cdot)$ is entire in $\C$ and
 for any $(x,\l)\in [0,\g]\ts\C,$ in addition to estimates in part 1), the following estimates ($\e :=\Im\l$) hold
 true:
\[
\lb{xn}
  |\c_n(x,\l )|\leq e^{(\g-x)(|\e|-\e)}\frac{1}{(2n)!}
  \left(\int_x^\gamma|q(\t)|d\t\right)^{2n},\ \ \forall \ n\geq 1,\]
\[
\lb{x}
  |\chi(x,\l)|\leq e^{(\g-x)(|\e|-\e)}\F(x),
\]
\[
\lb{x1} |\chi(x,\l)-1|\leq e^{(\g-x)(|\e|-\e)}\left(\F(x)-1\right).
\]

\end{lemma}
\no {\bf Proof.} The statements of part 1) of the Theorem and estimates (\ref{xn+}), (\ref{x+}), (\ref{x1+}) are well-known and can be found for example  in \cite{ZMNP} and \cite{DEGM}. We will not give any separate proof of these results, merely stating that these facts will  follow immediately by adapting our method of proving part 2) of the Theorem. Therefore we will first prove part 2)  under hypothesis that $q$ satisfy
Condition A and thereafter release this restriction while proving the estimates (\ref{x2}) and (\ref{L2norm}).

   Let $ t=(t_j)_1^{2n}\in \R^{2n}$ and $\mD_t(n)=\{x=t_0<t_1< t_2<...< t_{2n}<\g\}$.
Then using \er{series} we obtain
$$
\c_n(x,\l)=\int\limits_{\mD_t(n)} \lt(\prod\limits_{1\le j\le n}
q(t_{2j-1})\ol{q}(t_{2j}) e^{i2\l (t_{2j}-t_{2j-1})} \rt)dt, \qqq
t=(t_j)_1^{2n}\in \R^{2n},
$$
which yields
\[
\lb{y4}
\begin{aligned}
 &|\c_n(x,\l)|\le \int\limits_{\mD_t} \lt(\prod\limits_{1\le
j\le n}e^{(|\e|-\e)(t_{2j}-t_{2j-1})}
|q(t_{2j-1})q(t_{2j})|\rt)dt\\
&=\int\limits_{\mD_t(n)} \lt(\prod\limits_{1\le j\le {2n}}
|q(t_j)|\rt) e^{(|\e|-\e) \sum_1^n
(t_{2j}-t_{2j-1})}dt\\
&\le e^{ (\g-x) (|\e|-\e)}\int\limits_{\mD_t(n)}
|q(t_1)q(t_2)....q(t_{2n})| dt =
e^{(\g-x)(|\e|-\e)}\frac{1}{(2n)!}\left(\int_x^\g|q(\t)|d\t\right)^{2n},
\end{aligned}
\]
which yields \er{xn}.

This shows that the series \er{series} converge uniformly on
bounded subset of $\C$. Each term of this series is an entire
function. Hence it follows from Vitali's theorem that the sum is an entire function. Summing the majorants
we obtain estimates  \er{x} and \er{x1}.

In rest of the proof we do not suppose Condition A.

We show \er{x2}. Let $\eta=\Im \l>0$. Then \er{x0}  implies
\[
\lb{G1} |G(x,x',\l)|\le |q(x')|\int_x^{x'}
e^{-2\e(x'-\t)}|q(\t)|d\t\le {|q(x')|\,\|q\|_2\/(2\Im \l)^{1\/2}}.
\]
Substituting \er{G1}, \er{x1} into \er{x0} we obtain
$$
|\chi(x,\l)-1|\le\int_x^\g |G(x,x',\l)\c (x',\l )|dx'\le \int_x^\g
{|q(x')|\,\|q\|_2\/(2\Im \l)^{1\/2}}\F(x')dx'
$$
$$
\le {\F(x)\/|\Im \l|^{1\/2}}\|q\|_2\|q\|_1, \qqq \forall \ \Im \l>0,
$$
which yields \er{x2}.

Now, we will prove (\ref{L2norm}). For a fixed $x$ let $\langle
g(x,\cdot),h(x,\cdot)\rangle_{L^2}$ denote the  scalar product in
$L^2(\R, d\l)$ with respect to the second argument. In order to
prove  (\ref{L2norm}) we calculate and estimate
\[
\lb{norm} \int_\R|\chi(x,\l)-1|^2d\l=\langle \sum_{n \geq
1}\chi_n(x,\cdot), \sum_{m \geq 1}\chi_m(x,\cdot)\rangle_{L^2}.
\]

Let $\s(t)=\sum\limits_{1\le j\le n}(t_{2j}-t_{2j-1})$ and
$s=(s_j)_1^{2n}.$ We have
$$\begin{aligned} &\langle \c_n(x,\cdot),\c_m(x,\cdot)\rangle =
\int\limits_\R\int\limits_{\mD_t(n)} \lt(\prod\limits_{1\le j\le n}
q(t_{2j-1})\ol{q}(t_{2j})\rt) e^{i2\l \s(t)}
dt\\
&\cdot\int\limits_{\mD_s(m)} \lt(\prod\limits_{1\le j\le m}
\ol{q}(s_{2j-1})q(s_{2j})\rt) e^{-i2\l \s(s)} ds\,d\l,
\end{aligned}
$$
where in the previous definition of the domain $\mD_\cdot(\cdot)$ the constant $\g$ should be replaced with $\infty$ if $q$ does not have compact support.
Using that
$$ \int\limits_\R e^{i2\l \rt( \s(t)-\s(s)\rt)}d\l=4\pi
\d\rt(\s(t)-\s(s)\rt),
$$
where $\delta(\cdot)$ is the delta-function, we get
$$
\begin{aligned}
&\frac{1}{4\pi}\langle \c_n(x,\cdot),\c_m(x,\cdot)\rangle =\\
&\int\limits_{\begin{array}l \mD_t(n)\ts
\mD_s(m)\end{array}}\lt(\prod\limits_{1\le j\le n}
q(t_{2j-1})\ol{q}(t_{2j})\rt) \delta\left(\s(t)-\s(s)\right)
\lt(\prod\limits_{1\le j\le m} \ol{q}(s_{2j-1})q(s_{2j})\rt)
dsdt\\
&=\int\limits_{\begin{array}l \mD_t(n)\ts \mD_s(m-1)
\end{array}}\lt(\prod\limits_{1\le j\le n}
q(t_{2j-1})\ol{q}(t_{2j})\rt) \lt(\prod\limits_{1\le j\le m-1}
\ol{q}(s_{2j-1})q(s_{2j})\rt)\\
& \cdot \ol{q}(s_{2m-1})q\left(\s(t)-\sum\limits_{1\le j\le
m-1}(s_{2j}-s_{2j-1})+s_{2m-1}\right)
 ds_1ds_2...ds_{2m-1}dt.
\end{aligned}
$$
Now, we can estimate the right hand side using the H{\"o}lder inequality
$$
\left|\int\limits_{s_{2m-2}}^\infty \ol{q}(s_{2m-1})q \left(\s(t)-
\sum\limits_{1\le j\le m-1}(s_{2j}-s_{2j-1})+s_{2m-1}\right)
ds_{2m-1}\right| \leq\|q\|_2^2,
$$
and get
$$\begin{aligned} &\langle \c_n(x,\cdot),\c_m(x,\cdot)\rangle \leq\\
&4\pi\int\limits_{\mD_t(n)} |q(t_1)q(t_2)....q(t_{2n})| dt
\int\limits_{\mD_s(m-1)}
|q(s_1)q(s_2)....q(s_{2m-2})| ds_1...ds_{2m-2}\\
&\cdot \|q\|_2^2 =
\frac{4\pi}{(2n)!(2m-2)!}\left(\int_x^\infty |q(\t)|d\t\right)^{2n+2m-2}\|q\|_2^2.
\end{aligned}
$$
Then using (\ref{norm}) we get (\ref{L2norm}).
 \hfill\BBox
\vspace{1cm}

 As (\ref{5.11}) yields
\[\lb{a}
 a(\l )=\c (0,\l )=1+\int_0^\infty G(0,s,\l)\c (s,\l )ds,\] we get
$$
a(\l )=1+\sum _{n\geq 1}a_n(\l ),\qq a_1(\l )=\int_0^\infty
G(0,s,\l)ds,\ \dots, \  a_n(\l )=\c _n(0,\l ).
$$

Now, if $q$ satisfies Condition A we use (\ref{btilda}) and get
\[\lb{wb}
\wt{b}(\l)=\p_{2}^+(0,\l )=-i\int_0^\infty e^{-i\l (x-t)}\overline{q}(t)
\p_{1}^+(t,\l )dt=-i\int_0^\g e^{i2\l t}\overline{q}(t) \c (t,\l
)dt.
\]
 Note that if $q'\in L^1(\R),$ then by integration
by parts we get the following asymptotics:
 \[
 \lb{Dodd} a(\l)=1-\frac{1}{2i\l}\int_\R |q(t)|^2 dt+o(\l^{-1}),\qq
 \wt{b}(\l)=o(\l^{-1}),\qq \Im\l\geq 0,\,\,|\l|\to\iy
 \]
(which hold even without supposing Condition A, but under the weaker condition that $q\in L^1(\R)\cap L^2(\R),$ $q'\in L^1(\R),$  see \cite{DEGM}, p. 305). In Section \ref{s-Dodd improved} we show that for $\l\in\R$ and for $|\l|\rightarrow\infty$ one can
replace $o(\l^{-1})$ in (\ref{Dodd}) by ${\mathcal O}(\l^{-2})$ (see asymptotics (\ref{uniform_bound})).

We summarize the properties of functions $a,\wt{b}$  without supposing
$q,q'\in L^1(\R)$ in the following  lemma.
\begin{lemma}
\lb{l_ab}
Suppose $q\in L^1(\R)\cap L^2(\R).$ Then the following facts hold true:\\
1) The function $a(\l)$ is continuous  in the closed half-plane $\Im\l\geq 0$ and entire in the open half-plane $\Im\l>0.$ For  $\Im\l\geq 0$ the function $a$ satisfies the following estimates:\\
\[
\begin{aligned}
\lb{estimate+}
& |a(\l)|\leq \ch \|q\|_1,\qq  |a(\l )-1|\leq \ch \|q\|_1 -1.
\end{aligned}
\]
For all $\Im \l>0,$
$$|a(\l )-1|\leq {\ch \|q\|_1\/|\Im \l|^{1\/2}}\|q\|_2\|q\|_1.$$

 Moreover,
\[\lb{a_in_L2} a(\cdot)-1\in L^2(\R).
\]

2) If $q$ satisfies Condition A, then for each $x\in \R $ the
functions  $\p^{\pm}(x,\l ), \vp^{\pm}(x,\l )$ and $a(\l),$ $b(\l),$
$\wt{b}(\l)$ are entire on $\C$. In addition to estimates in part 1),
 the following estimates hold true:
\[
\begin{aligned}
\lb{estimate}
& |a(\l)|\leq e^{\g(|\e|-\e)}\ch \|q\|_1,\\
& |a(\l )-1|\leq e^{ \g(|\e |-\e )}(\ch \|q\|_1 -1),\\
&\left| \wt{b}(\l)+i\int_0^\g e^{i2\l t}
\overline{q}(t)dt\right|\leq e^{\g (|\eta|-\eta)}\left(\sh \|q\|_1
-\|q\|_1\right),
\end{aligned}
\]
where  $\eta=\Im\l$.
\end{lemma}
{\bf Proof.} The results in Part 1) follows  directly from Lemma \ref{chi_estimates}  and formula (\ref{a}). The fact that $a(\cdot)-1\in L^2(\R),$ (\ref{a_in_L2}), follows from (\ref{L2norm}).

Suppose  that $q$ satisfies Condition A. Then representing $\chi(x,t)$ as a sum as in (\ref{series}),  estimating each term in the sum as in the proof of Lemma \ref{chi_estimates} , bounds (\ref{y4}), and by integrating by parts, we get
 $$\begin{aligned}
&\left| \wt{b}(\l)+i\int_0^\g e^{i2\l t}
\overline{q}(t)dt\right|\leq e^{\g (|\e |-\e )}\sum_{n\geq 1}\frac{1}{(2n)!}\int_0^\gamma |q(x)| \left(\int_x^\g|q(\t)|d\t\right)^{2n}dx=\\
&=e^{\g (|\e |-\e )}\sum_{n\geq 1}\frac{1}{(2n+1)!} \left(\int_x^\g|q(\t)|d\t\right)^{2n+1}dx=e^{\g(|\eta|-\eta)}\left(\sinh \|q\|_1 -\|q\|_1\right)
\end{aligned},$$
which shows the last inequality in (\ref{estimate}).

\hfill\BBox

If $q$ satisfies Condition A, then using the analyticity of $a,b,\wt{b},$ identity
(\ref{identity-on-real}) has an analytic  continuation into the whole complex plane  as
\begin{equation}\lb{identity-on-complex}
a(\l)\overline{a}(\overline{\l})-b(\l)\overline{b}(\overline{\l})=1,\qq\l\in\C.
\end{equation}

 All zeros of $a(\l, q)$   lie in $\C_-.$
Denote by $\{\l_n\}_{1}^{\iy}$ the sequence of its zeros in $\C_-$
(multiplicities counted by repetition), so arranged that
$0<|\l_1|\leq |\l_2|\leq |\l_2|\leq \dots.$ We denote the number of
zeros of function $a$ having modulus  $\leq r$ by $\cN (r,a)$, each
zero being counted according to its multiplicity.

We will need the following Lemma by Froese (see  \cite{F97}, Lemma 4.1). Even though the original lemma was stated for $V\in L^\infty$ the argument also works for $V\in L^2$ and we reproduce this version of lemma here for the sake of completeness.
\begin{lemma}\lb{l-Fr}  Suppose $V\in L^2(\R)$ has compact support
contained in $[0,1],$ but in no smaller interval. Suppose $f(x,\l)$
is analytic for $\l$ in the lower half plane, and for real $\l$ we
have $f(x,\l)\in L^2([0,1]\,dx,\R\, d \l).$ Then $\int_\R e^{i\l x}
V(1-f(x,\l))\,d x$ has exponential type at least $1$ for $\l$ in the
lower half plane.\end{lemma}

In the following Proposition we state the analytic properties of functions $a,\wt{b}.$

\begin{proposition}
\lb{Prop2}
Assume that potential $q$ satisfies
Condition A. Then
\[
a(\cdot) \in \cE_1(2\g),\qqq \wt{b}(\cdot)\in \cE_0(2\g),
\]
\[\begin{aligned}\lb{as-im}
&a(i\e ,q)=1+ o(1),\qq \wt{b}(i\e,q)=-i\int_0^\g e^{-2\e t}
\overline{q}(t)dt+o(1)\qq\mbox{as}\qq\eta\rightarrow\infty,
\end{aligned}\]
   and
\[\lb{Had-fact}
a(\l ,q)=a(0,q)e^{i\g\,\l }\lim_{r\to +\infty} \prod_{|z_n|\le r}
\lt(1-{\l \/ \l _n} \rt), \ \ \ \l \in \C ,
\]
uniformly in every disc.

\end{proposition}
{\bf Proof.} First we prove that $a(\cdot) \in \cE_1(2\g),$
$\wt{b}(\cdot)\in \cE_0(2\g).$ By (\ref{estimate}), functions $a,$ $\wt{b}$
have exponential type in the lower half plane at most $2\g:$
$\rho_-(\wt{b})\leq 2\g.$ Now, we have by (\ref{x1})
$$\wt{b}(\l)=-i\int _0^1 e^{i2\l t}\overline{q}(t) \c (t,\l
)dt=-i\int _0^1 e^{i2\l t}\overline{q}(t) (1+X (t,\l ))dt,$$ where
$X(t,\l)=\chi(t,\l)-1$ is analytic in $\C_-$ and $\int_0^\g
dx\int_\R d\l |X(x,\l)|^2 <\infty$ by Lemma \ref{chi_estimates},
bound (\ref{L2norm}). Using that the support of $q$ is contained in
$[0,1],$ but in no smaller   interval (Condition A), we get that
 $\wt{b}(\l)$ has exponential type of at least $\rho_-=2\g$ by
Lemma \ref{l-Fr}.

 The proof of $\rho_+=0$ is similar.
Now, using (\ref{identity-on-complex}),
$a(\l)\overline{a}(\overline{\l})=1+b(\l)\overline{b}(\overline{\l}),$
and $\wt{b}(\l)=-\overline{b}(\overline{\l}),$ we get the same
result for the function $a(\l).$ The asymptotics (\ref{as-im})
follows from (\ref{estimate}).

 Inequality $\int_\R\frac{\log(1+|f(\l)|)}{1+\l^2}d\l <\infty,$ where $f=a(\l)$ or $f=\wt{b}(\l),$  follows trivially  from the fact that $a,\wt{b}\in L^\infty(\R).$
From (\ref{identity-on-real}) it follows that $|a(\l)|\geq 1$ for $\forall\,\l\in\R.$ Therefore we have $a(\cdot) \in \cE_1(2\g)$ and
$\wt{b}(\cdot)\in \cE_0(2\g).$

 Formulas in (\ref{as-im}) follow from  bounds (\ref{estimate}) respectively (\ref{x2}).

Formula (\ref{Had-fact}) is the standard Hadamard factorization of a function from Cartwright class, see (\ref{Hadfact}).
 \hfill$\BBox$

{\bf Proof of Theorem \ref{T2}.}
Proposition \ref{Prop2} shows that the conditions of  Levinson Theorem \ref{th-Levinson} are fulfilled which  gives  asymptotics (\ref{counting}).\hfill\BBox

\section{Estimates for $\psi$ for the case  $q'\in L^1(\R).$}\lb{s-Dodd improved}
\setcounter{equation}{0}
We suppose that $q$ satisfies Condition A and $q'\in L^1(\R).$
We consider the Dirac equation \[\lb{Direq}-i\sigma_3\psi'+V\psi=\l\psi,\qq V=\left(
                                                                                \begin{array}{cc}
                                                                                  0 & q \\
                                                                                  \overline{q} & 0 \\
                                                                                \end{array}
                                                                              \right),
\] where we use the Pauli notation $J=\sigma_3.$ Note the following commutation properties
\[\lb{commut}\sigma _3V=-V\sigma_3,\,\, e^{i\l t\sigma_3}V=e^{-i\l t\sigma_3}V,\,\,\sigma_3^2=I_2.
\] Recall that the Jost solution $\psi^+=(\psi_1^+,\psi_2^+)^{\rm T}$ is solution of (\ref{Direq}) satisfying the condition
\[\lb{Direqcond} \psi^+=e^{i\l x}e_+\equiv e^{i\l x\sigma_3}\left(
                                                        \begin{array}{c}
                                                          1 \\
                                                          0 \\
                                                        \end{array}
                                                      \right),\qq\mbox{for}\,\, x>\gamma.\]

The main result of this section is the following Lemma
\begin{lemma}\lb{l-smooth q} Let $q$ satisfy Condition A and $ q'\in L^1(\R).$ Then for $|\l|\geq \sup_{t\in\R}|q|$ the Jost solution $\psi=\psi^+$ of  (\ref{Direq}),
(\ref{Direqcond}) satisfies
$$
\begin{aligned}
&\psi(x,\l)=\psi^0+\frac{1}{2\l}a^{-1}K\psi,\qq \psi^0=a^{-1}e^{i\l x\sigma_3}\left(
                      \begin{array}{c}
                        1 \\
                        0 \\
                      \end{array}
                    \right), \qq KY=\int_x^\gamma
e^{i\sigma_3(x-t)} W(t)\psi(t,\l)dt,\\
&a(x,\l)=I_2-\frac{1}{2\l}V(x),\qq W(t)=V'(t)+i| q(t)|^2\sigma_3,\\
& \psi=\psi^0+\sum_{n\geq 1}\psi^n,\qq  \psi^n=\frac{1}{(2\l)^n}(a^{-1}K)^n\psi^0,
\end{aligned}$$
where the series converge uniformly on bounded subsets of $\{ (x,\l);\,\,x\in\R, |\l|\geq \sup_{t\in\R}|q|\}$ and for any $j\geq 2,$ the following estimates hold true:
 $$|\psi^n(x,\l)|\leq\frac{2}{n!|\l|^n}e^{|\Im \l|(2\gamma-x)}\left(\int_0^\gamma |W(s)|ds\right)^n.  $$

 Let  $a(\l)=\psi_1^+(0,\l).$ Then for any $|\l|\geq \sup_{t\in\R}|q|$
\[\lb{uniform Dodd}
a(\l)=1-\frac{1}{2i\l}\|q\|_2^2 +{\mathcal O}\left(\frac{e^{|\Im\l| 2\gamma}}{|\l|^2}\right).
\]
Moreover, the quantity \[\lb{uniform_bound}
\sup_{\l\in\R}\left|
\l^2\left(a(\l)-1+\frac{1}{2i\l}\|q\|_2^2\right)\right|
\] is finite.
\end{lemma}
{\bf Proof.} We use the arguments from \cite{K08}.
Note that (\ref{Direq}) is equivalent to\\
$\psi'-i\l\sigma_3\psi=-i\sigma_3 V\psi$ and
\[\lb{Direq2} \left(e^{-i\l x\sigma_3}\psi\right)'=-ie^{-i\l x\sigma_3}\sigma_3 V\psi.\]
The Jost function $\psi=\psi^+$ satisfies the integral equation
$$
 \p(x,\l )=e^{i\l x\sigma_3}e_1+\int _x^\g i\sigma_3 e^{i\l
(x-t)\sigma_3} V(t) \p(t,\l )dt.
$$
Using (\ref{commut}) we write it in the form
$$
 \p(x,\l )=e^{i\l x\sigma_3}e_1+\int _x^\g i\sigma_3 e^{i\l
(x-2t)\sigma_3} V(t)e^{-i\l t\sigma_3} \p(t,\l )dt.
$$
Using that $q'\in L^1$ we integrate by parts and use that $e^{-i\l t\sigma_3} \p(t,\l )$ satisfies (\ref{Direq2})
$$\begin{aligned}
\psi(x,\l)=&e^{i\l x\sigma_3}e_1+\left[\frac{i\sigma_3^2}{-i2\l}e^{i\l(x-2t)\sigma_3} V(t)e^{-i\l t\sigma_3} \p(t,\l )\right]_{t=x}^\gamma -\\
&-\int_x^\gamma \frac{i\sigma_3^2}{-i2\l}e^{i\l(x-2t)\sigma_3}\left(V'(t)e^{-i\l t\sigma_3} -iVe^{-i\l t\sigma_3}\sigma_3 V\right)\p(t,\l ).
\end{aligned}
$$
Again using the commutation relations (\ref{commut}) we get the integral equation
$$\psi(x,\l)=e^{i\l x\sigma_3}e_1+\frac{1}{2\l} V(x)\psi(x,\l)+\frac{1}{2\l}\int_x^\gamma e^{i\l (x-t)\sigma_3}\left(V'+i| q|^2\sigma_3\right)\psi(t,\l)dt.$$
Put $W(t)=V'+i| q|^2\sigma_3$ and $a(x,\l)=I_2-\frac{1}{2\l}V(x).$ Then $\psi$ satisfies
$$a(x,\l)\psi(x,\l)=e^{i\l x\sigma_3}e_1+\frac{1}{2\l}\int_x^\gamma e^{i\l (x-t)\sigma_3}W(t)\psi(t,\l)dt.$$
Using that  \[\lb{ba}
\mbox{for}\qq
|\l|\geq \sup_{t\in\R}|q|\qq\mbox{we have}\qq \sup_{t\in\R}|a^{-1}|\leq 2,\]  we get the integral equation
$$\psi(x,\l)=\psi^0+\frac{1}{2\l} a^{-1}K\psi,\qq \psi^0=a^{-1}e^{i\l x\sigma_3}\left(
                      \begin{array}{c}
                        1 \\
                        0 \\
                      \end{array}
                    \right), \qq K\psi=\int_x^\gamma
e^{i\sigma_3(x-t)} W(t)\psi(t,\l)dt.$$   By iterating we get
$$\psi=\psi^0+\sum_{n\geq 1}\psi^n,\qq  \psi^n=\frac{1}{(2\l)^n}(a^{-1}K)^n\psi^0.$$
Let $ t=(t_j)_1^{n}\in \R^{n}$ and $\mD_t(n)=\{x=t_0<t_1< t_2<...< t_{n}<\gamma\}$.
$$\psi^n=\frac{1}{(2\l)^n}\int_{\mD_t(n)}\prod_{j=1}^n (a(t_{j-1}))^{-1}e^{i\l \sigma_3 (t_{j-1}-t_j)}W(t_j)(a(t_n))^{-1}e^{i\l t_n\sigma_3}\left(
                      \begin{array}{c}
                        1 \\
                        0 \\
                      \end{array}
                    \right)dt.$$
                                     Let $x>0.$  Now, using
 (\ref{ba}) and
 $$\left|e^{ i\l \sigma_3 (t_{j-1}-t_j)}\right|\leq e^{|\Im\l| (t_j-t_{j-1})},\qq \sum_{j=1}^n(t_j-t_{j-1}) =t_n-t_0,\qq \left| e^{i\l t_n\sigma_3}\right| \leq
 e^{|\Im\l| t_n},$$
  we get
 $$|\psi^n(x,\l)|\leq \frac{2}{|\l|^n}e^{|\Im \l|(2\gamma-x)}\int_{\mD_t(n)}\prod_{j=1}^n | W(t_j)|dt=\frac{2}{n!|\l|^n}e^{|\Im \l|(2\gamma-x)}\left(\int_0^\gamma |W(s)|ds\right)^n.  $$

 Note that explicitly $$a^{-1}=\frac{1}{1-(2\l)^{-2}|q|^2}\left(
                                             \begin{array}{cc}
                                               1 & (2\l)^{-1}q \\
                                               (2\l)^{-1}\overline{q} & 1 \\
                                             \end{array}
                                           \right),\qq W(t)=\left(
                                                              \begin{array}{cc}
                                                                i|q|^2 & q' \\
                                                                \overline{q}' & -i|q|^2 \\
                                                              \end{array}
                                                            \right)
                                           $$
and
$$\psi^0=\frac{1}{1-(2\l)^{-2}|q|^2}e^{i\l x}\left(
                      \begin{array}{c}
                        1 \\
                        0 \\
                      \end{array}
                    \right).$$
 Putting $x=0$ we get
 $$\psi^0_1(0,\l)=1+{\mathcal O}(\l^{-2}),\qq \psi^1_1(0,\l)=-\frac{1}{2i\l}\int_0^\gamma|q|^2dt +{\mathcal O}(\l^{-2})$$ and as $a(\l)=\psi_1^+(0,\l)$ we get
(\ref{uniform Dodd}).

Now, for $\l\in\R$   bound (\ref{uniform Dodd})  implies that       (\ref{uniform_bound}),
 which is used in Theorem \ref{T3}. \hfill\BBox

\section{The resolvent estimates }\lb{s-determinant}
\setcounter{equation}{0}


Let $R(\l)=(H-\l I)^{-1}$ denote resolvent for operator $H$ and let $R_0(\l)$ be the free resolvent (i.e. for the case $q= 0$).
We have
$$
R_0(\l)=(H_0-\l I)^{-1}=\left(
                            \begin{array}{cc}
                              (-i\pa_x-\l)^{-1} & 0 \\
                              0 & (i\pa_x-\l)^{-1} \\
                            \end{array}
                          \right)=\left(
                            \begin{array}{cc}
                              T_0 & 0 \\
                              0 & S_0 \\
                            \end{array}
                          \right),
$$
where $\pa_x={d\/dx},$ and
$$
T_0f(x)= i\int_{-\infty}^xe^{i\l(x-y)}f(y)dy,\qq S_0f(x)=
                     i\int_{x}^\infty e^{-i\l(x-y)}f(y)dy\qq\mbox{if}\,\,\Im\l >0;$$
$$ T_0f(x)= -i\int_{x}^\infty e^{i\l(x-y)}f(y)dy, \qq
S_0 f(x)= -i\int_{-\iy}^xe^{-i\l(x-y)}f(y)dy\qq\mbox{if}\,\,\Im\l
<0.
$$

We denote by $\|.\|_{\cB_k},$ the Trace ($k=1$) and the Hilbert-Schmidt ($k=2$) operator norms.

For a Banach space $\cX,$ let $AC(\C_\pm;\cX)$ denote the set of all
$\cX$-valued continuous functions on
$\overline{\C}_\pm=\{\l\in\C;\,\,\pm\Im \l\geq 0\},$ which are analytic
on $\C_\pm$.

\begin{lemma}
\lb{L-HS} Let $\r,\wt\r\in L^2(\R;\C^2)$.  Then it follows:\\
i) Operators $\r R_0(\l), R_0(\l)\r, \r R_0(\l) \wt\r$ are the $\cB_2$-valued
operator-functions satisfying the following properties:
\[
\begin{aligned}
\lb{B2-1} \|\rho R_0(\l)\|_{\cB_2}^2=\| R_0(\l)\rho\|_{\cB_2}^2 =
\frac{\|\rho\|_2^2}{2|\Im\l|},
\end{aligned}
\]
\[
\lb{B2-2} \|\r R_0(\l)\wt\r \|_{\cB_2}\le \|\r\|_2\|\wt\r\|_2,\qq \|\r R_0(\l)\wt\r \|_{\cB_2}\rightarrow 0\,\,\mbox{as}\,\,|\Im\l|\rightarrow\infty.
\]
Moreover, the operator-function $\r R_0\wt\r\in AC(\C_\pm; \cB_2)$.

ii) Operator $ \r R_0'(\l) \wt\r=\r R_0^2(\l) \wt\r$ is the $\cB_2$-valued
operator-functions,  analytic in $\C_\pm:$\\ $\r R_0'\wt\r\in AC(\C_\pm; \cB_2),$ satisfying
\[
\lb{B2-3}
\|\r R_0'(\l)\wt\r \|_{\cB_2}\le \frac{\|\r\|_2\|\wt\r\|_2}{e|\Im\l|},\qq \|\r R_0'(\l)\wt\r \|_{\cB_2}\rightarrow 0\,\,\mbox{as}\,\,|\Im\l|\rightarrow\infty.\]

\end{lemma}
{\bf Proof.} i) Let $\Im\l \ne 0$ and $\chi\in L^2(\R;\C).$  Then the Fourier transformation
implies
$$
\|\chi (\mp i\pa_x-\l)^{-1}\|_{\cB_2}^2={1\/2\pi}\int_\R |\chi(x)|^2dx
\int_\R {dk\/|\pm k-\l|^2} ={1\/2|\Im\l|}\|\chi\|_2^2.$$
As
$$\begin{aligned}\|\rho R_0(\l)\|_{\cB_2}^2=&\Tr ( (\rho R_0(\l))^*\rho R_0(\l))\\
=&\|\rho_{11} T_0(\l)\|_{\cB_2}^2+\|\rho_{21} T_0(\l)\|_{\cB_2}^2+\| \rho_{12}S_0(\l)\|_{\cB_2}^2+\| \rho_{22}S_0(\l)\|_{\cB_2}^2\end{aligned}$$ we get the estimate. The proof for $R_0(\l)\r$ is similar.
This yields identity \er{B2-1}. This  identity and the resolvent
identity
\[
\lb{RI}
\r R_0(\l)=\r R_0(\m)+ \ve\r  R_0^2(\m)+\ve^2 \r
R_0(\l)R_0^2(\m), \qqq \ve=\l-\m,
\]
yields that the mapping $\l\to \r R_0(\l)$ acting from $\C_+$
into $\cB_2$ is analytic.

Let $\Im\l >0$ (for $\Im\l <0$ the proof is similar) and $\chi,\wt{\chi}\in L^2(\R;\C).$  Then
  we have
$$
\|\chi T_0(\l)\wt\chi\|_{\cB_2}^2=\int_{\R} |\chi(x)|^2\int_{-\iy}^x
e^{-2(x-y)\Im\l}|\wt\chi(y)|^2dy dx \le \|\chi\|_2^2\|\wt\chi\|_2^2,
$$
and similar we get $\|\chi S_0\wt\chi\|_{\cB_2}^2\le
\|\chi\|_2^2\|\wt\chi\|_2^2.$ These bounds and the dominated convergence Lebesgue Theorem yields  (\ref{B2-2}).

 Moreover, these arguments show that each
$\r R_{0}(\l\pm i0)\wt\r\in \cB_2, \l\in \R$.

We show that $\r R_0\wt\r\in AC(\C_\pm, \cB_2)$. Let $\l,\m\in
\ol\C_+ $ and $\m\to \l$ We have
$$
\|\chi (T_0(\l)-T_0(\m))\wt\chi\|_{\cB_2}^2=\int_{\R}
|\chi(x)|^2\int_{-\iy}^x X(x,y,\l,\m) |\wt\chi(y)|^2dy dx, \qq
$$
where the function $X(x,y,\l,\m)=|e^{i(x-y)\l}-e^{i(x-y)\m}|^2, x>y$
satisfies
$$
X(x,y,\l,\m)\le 2,\qqq  {\rm and }\qq X(x,y,\l,\m)\to X(x,y,\l,\l) \
\as \ \m\to \l,\ \ \forall x>y.
$$
Writing the similar identity for $S_0$ and applying Lebesgue Theorem yields that the operator-function $\r
R_0\wt\r\in AC(\C_\pm, \cB_2)$.

ii) The proof of  (\ref{B2-3}) is easily verified as in the proof of i).

It is sufficient to prove for $\Im\l >0$ and $\chi,\wt{\chi}\in L^2(\R;\C).$  Then
  we get
$$
\|\chi T_0'(\l)\wt\chi\|_{\cB_2}^2=\int_{\R} |\chi(x)|^2\int_{-\iy}^x
(x-y)^2e^{-2(x-y)\Im\l}|\wt\chi(y)|^2dy dx \le \frac{\|\chi\|_2^2\|\wt\chi\|_2^2}{e^2|\Im\l|^2},
$$
where we used that the function $t^2e^{-2t\Im\l}\leq (e|\Im\l|)^{-2}$ for $t\geq 0.$
Similar we get $\|\chi S_0'\wt\chi\|_{\cB_2}^2\le  (e|\Im\l|)^{-2}
\|\chi\|_2^2\|\wt\chi\|_2^2.$ These bounds and the dominated convergence Lebesgue Theorem yields  (\ref{B2-3}) and as in i) we get that
$\r
R_0'\wt\r\in AC(\C_\pm, \cB_2).$ \hfill $\BBox$

We pass now to study of the full resolvent $R(\l)=(H_0+V-\l I)^{-1}.$  We factorize $V$ as follows
$$
V=\ma 0&q\\ \ol q & 0\am=V_1V_2, \qqq {\rm  where}\qqq  V_2=|q|^{1\/2}I_2.
$$

In the beginning we do not suppose that $q$ satisfies Condition A, but just $q\in L^2(\R).$

Let $Y_0(\l)=V_2R_0(\l)V_1,$ $Y(\l)=V_2R(\l)V_1.$ Then we have
\begin{equation}\lb{res-id}
Y(\l)=Y_0(\l)-Y_0(\l)\left[I+Y_0(\l)\right]^{-1}Y_0(\l),\qq Y=I-(1+Y_0)^{-1}\end{equation} and
\begin{equation}\lb{2.5}
 (I+Y_0(\l))(I-Y(\l))=I.
\end{equation}

\begin{corollary}
\lb{Ttrace}
Let $q\in L^2(\R)$ and let $\Im\l\neq 0$. Then\\
\no i)
\[
\lb{tr1}\|VR_0(\l)\|_{\cB_2}^2= {\|q\|_2^2\/|\Im\l|},
\]
\no ii) The operator $R(\l)-R_0(\l)$ is of trace class and satisfies
\[
\lb{tr2}\|R(\l)-R_0(\l)\|_{\cB_1}\leq \frac {C}{|\Im\l|},
\] for some constant $C.$

iii) Let, in addition, $q\in L^1(\R)\cap L^2(\R)$. Then
 we have $Y_0, Y, Y_0', Y'\in AC(\C_\pm; \cB_2)$ and the following following bounds are
 satisfied:
\[
\lb{eY0} \|Y_0(\l)\|_{\cB_2}\le \|q\|_1,\qq \forall \ \l\in \C;\qqq \|Y_0(\l)\|_{\cB_2}\rightarrow 0\qq\mbox{as}\,\,|\Im\l|\rightarrow\infty.
\]
\[
\lb{eY0'} \|Y_0'(\l)\|_{\cB_2}\le \frac{\|q\|_1}{e|\Im\l|},\qq \forall \ \l\in \C\setminus\R;\qqq \|Y_0'(\l)\|_{\cB_2}\rightarrow 0\qq\mbox{as}\,\,|\Im\l|\rightarrow\infty.
\]

\end{corollary}
{\bf Proof.}
i) Identity (\ref{tr1}) follows from (\ref{B2-1}) and $\|V\|_2^2=2\|q\|_2^2,$ but also directly from:
$$ \|VR_0(\l)\|_{\cB_2}^2=\Tr ( V R_0(\l))^*V R_0(\l))=\Tr (\overline{q}T_0)^*\overline{q}T_0 +\Tr (qS_0)^*qS_0=\|\overline{q}T_0\|_{\cB_2}^2+\|qS_0\|_{\cB_2}^2.  $$
ii)  Denote $\cJ_0(\l)=I+Y_0(\l).$ For $\Im\l\neq 0,$ operator $\cJ_0(\l)$ has bounded inverse and the operator
 $$
R(\l)-R_0(\l)=-R_0(\l)V_1\left[\cJ_0(\l)\right]^{-1}V_2R_0(\l),\qq \Im\l\neq 0,
$$
is trace class and  the estimate follows from (\ref{tr1}).

iii)  That $Y_0, Y\in AC(\C_\pm; \cB_2)$ follows as in the proof of Lemma \ref{L-HS}, resolvent identity (\ref{res-id}) and ii), bound (\ref{tr2}).  Using (\ref{2.5}), we get
$$Y'(\l)=(I-Y(\l))Y_0'(\l)(I-Y(\l))\in AC(\C_+;\cB_2).$$

 We put $\Im\l >0$ (for $\Im\l <0$ the proof is similar).
  The first inequality in   (\ref{eY0}) follows as (\ref{B2-2}) in Lemma \ref{L-HS} using the off-diagonal form of matrix-function $Y_0$:  \[\lb{above}
\begin{aligned}
 \|Y_0(\l)\|_{\cB_2}^2&= \Tr(Y_0^*(\l)Y_0(\l))=\Tr(T_0|q|T_0^*|q|) +
\Tr(S_0|q|S_0^*|q|)\\&=2\int_0^\infty\int_0^x
e^{-2\Im\l(x-t)}|q(t)|dt|q(x)|dx
 \leq\left(\int_0^\infty |q(x)|dx\right)^2.\end{aligned}
 \]
By the dominated convergence (Lebesgue) Theorem this also shows that $\Tr(Y_0^*(\l)Y_0(\l))\rightarrow 0$ as $\Im\l\rightarrow\infty,$ proving the second property in (\ref{eY0}).

The properties (\ref{eY0'}) follows as in (\ref{B2-2}) in Lemma \ref{L-HS} and similarly to (\ref{above}) by using the simple form of $Y_0'$ .

\hfill $\BBox$

\begin{lemma}
\lb{TY} Let $q\in  L^2(\R)$. Then
 \[
\lb{2.9} VR_0^2(\l),\,\,Y_0'(\l)\in \cB_1,\qq \Im\l\neq 0.
\]
\[
\lb{2.10} \Tr Y_0'(\l)=0,\qq \Tr Y_0^n(\l) =0,\qq\Im\l\neq 0,\qq \forall\,  n\in 2\N+1. \
\]
\[
\lb{l-trace0} \Tr(Y_0(\l+i0)-Y_0(\l-i0))=0,\qq\forall\,
\l\in\R.
\]

\end{lemma}
{\bf Proof.} We prove (\ref{2.9}). We have $VR_0^2(\l)\in \cB_1$ by recalling
that  $R_0(\l)={\rm
diag}\,((-i\partial_x-\l)^{-1},(i\partial_x-\l)^{-1})$ and applying
Theorem XI.21 in \cite{RS-vIII}, stating that
$$f(x)g(-i\partial_x)\in\cB_1,\,\,f,g\in L^{2,\delta}(\R^n),\,\,\delta >n/2,$$ where $f\in L^{2,\delta}(\R^n)$ means $\int_{\R^n}(1+|x|)^{2\delta}|f(x)|^2dx <\infty.$

 As $Y_0'(\l)=V_2R_0^2(\l)V_1,$   $\|Y_0'\|_{\cB_1}\leq\|V_2R_0\|_{\cB_2}\cdot\| R_0V_1\|_{\cB_2},$ and applying Lemma \ref{L-HS} we get that $Y_0'(\l)$
is trace class.

 The first identity in (\ref{2.10}) follows from the identities $\Tr Y_0'(\l)=\Tr
VR_0^2(\l)=0$ as
$$
VR_0'(\l)= \ma 0&q\\ \ol{q}& 0\am \left(\begin{array}{cc}
                              T_{0}'(\l) & 0 \\
                              0 & S_{0}'(\l) \\
                            \end{array}
                          \right)=\ma 0&q S_{0}'\\ \ol{q}T_{0}'& 0\am
$$
is off-diagonal matrix operator.

The second identity in (\ref{2.10}) follows as $\Tr Y_0^n(\l) =\Tr \rt(V R_0(\l)\rt)^n$ and $\left(V R_0(\l)\right)^n$ is  off-diagonal matrix operator for $n$ odd.

 Formula (\ref{l-trace0} ) follows similarly as
$$
 \rt[R_0(\l+i0)-R_0(\l-i0)\rt](x,y)=i\left(
                         \begin{array}{cc}
                           e^{i\l(x-y)} & 0 \\
                           0 & e^{-i\l(x-y)} \\
                         \end{array}
                       \right),
 $$
 and the product $V( R_0(\l+i0)-R_0(\l-i0))$ is  off-diagonal
 matrix-valued operator.
 \hfill\BBox

\section{Modified Fredholm determinant}\lb{s-ModFrDet} In this section we will follow our agreement that the "physical sheet" corresponds to $\C_+$ and the resonances lie in $\C_-$ (see Introduction).

The "sandwiched" resolvent $Y_0(\l):=V_2R_0(\l)V_1\in \cB_2$ is not trace class as the integral kernel of $R_0(\l)$ in the Fourier representation has non-integrable singularities $(\pm k-\l)^{-1}.$ However, it was shown in Corollary \ref{Ttrace} that $Y_0(\l)$
 is Hilbert-Schmidt, and
 we define the modified Fredholm determinant
$$D(\l)=\det\left[ (I+Y_0(\l)) e^{-Y_0(\l)}\right],\qq\l\in\C_+.$$

\begin{lemma}
\lb{TD1} Let $q\in L^1(\R)\cap L^2(\R)$. Then

i) The function $D$ belongs to $AC(\C_+;\C)$ and satisfies:
\[
\lb{ED1}
D'(\l)=-D(\l)\Tr\left[Y(\l)Y_0'(\l)\right]\qq\forall\l\in\C_+,
\]
\[
\lb{ED2} |D(\l)|\le e^{\frac12\|q\|_1^2},\qq\forall\l\in\C_+,
\]
\[
\lb{ED3} D(\l)\ne 0, \qqq \qq\forall\l\in\ol\C_+,
\]
\[\lb{ED10} D(\l)\rightarrow 1\qq\mbox{as}\qq\Im\l\rightarrow\infty.
\]
ii) The functions $\log D(\l)$ and $\frac{d}{d\l}\log D(\l)$ belong
to $AC(\C_+;\C),$ and the following identities  hold:
\[
\begin{aligned}
\lb{SD} &-\log D(\l)=\sum_{k=
1}^\infty{\Tr Y_0^{2k}(\l)\/2k}=\sum_{k=1}^\infty\frac{\Tr \left( T_0(\l)qS_0(\l)\overline{q}\right)^k}{k},
\end{aligned}
\]
 where the series converge absolutely and uniformly for
 $|\Im \l|>2\| q\|_2^2=\| V\|_2^2,\,\,\l\in\C_+,$
  and
\[
\lb{SD2} \rt|\log D(\l)+\sum_{n=1}^{N}{\Tr Y_0^{2n}(\l)\/2n}\rt|\le
{\ve_\l^{N+1}\/(N+1)(1-\ve_\l)},\,\,\ve_\l={\|V\|_2^2\/2|\Im
\l|},\,\,\l\in\C_+,
\]
for any $N\ge 1.$ Moreover, $\frac{d^k}{d\l^k}\log D(\l)\in AC
(\C_+;\C)$ for any $k\in\N.$
\end{lemma}
\no {\bf Proof.} i) Formula (\ref{ED1}) is well-known (see for example
\cite{GK69}) and together with iii) in Corollary
\ref{Ttrace}  it implies  that the
functions $\log D(\l)$ and $\frac{d}{d\l}\log D(\l)$ belong to
$AC(\C_+;\C).$ Estimate (\ref{ED2}) follows from the
inequality ((2.2), page 212, in russian edition of \cite {GK69})
\begin{equation}\lb{GK}
|D(\l)|\leq e^{\frac12\Tr(Y_0^*(\l)Y_0(\l))}
\end{equation} and inequality (\ref{eY0}): $\Tr(Y_0^*(\l)Y_0(\l))\leq \|q\|_1^2.$
As the zeros of $D(\l)$ in $\C_+$ are the eigenvalues of $H$ and $H$ does not have eigenvalues it follows  (\ref{ED3}).

 Property (\ref{ED10}) will follow from estimate (\ref{SD2}) in the part ii) of Lemma. Below we will prove it.

ii) Denote $F(\l)=\sum_{n\ge 2}{\Tr (-Y_0(\l))^n\/n}.$ By (\ref{2.10}) in Lemma \ref{TY}, $F(\l)$ coincides with the series in \er{SD}. We show that this series
converge absolutely and uniformly. Indeed, as in the proof of \er{tr1} we get
\[\lb{trn} \left| \Tr \left( T_0(\l)qS_0(\l)\overline{q}\right)^k\right| \leq\| T_0(\l)q\|_{\cB_2}^k\cdot \| S_0(\l)\overline{q}\|_{\cB_2}^k=\ve_\l^k,
\] where $$\ve_\l=\| T_0q\|_{\cB_2}^2=\| S_0\overline{q}\|_{\cB_2}^2=\frac{\|q\|_2^2}{|\Im\l|}=\frac{\|V\|_2^2}{2|\Im\l |}.$$

Then $F(\l)$  is analytic function in the domain $|\Im \l|>
\|V\|_2^2$. Moreover, by differentiating $F$ and using (\ref{2.5}) we get
$$
F'(\l)=-\lim_{m\to\iy}\sum_{n\ge 2}^m\Tr (-Y_0(\l))^{n-1}Y_0'(\l)=
\Tr Y(\l)Y_0'(\l),  \ \ \ |\Im \l|> \|V\|_2^2,
$$
and then the function $F=\log D(\l)$, since $F(i\tau)=o(1)$ as $\tau\rightarrow\infty.$ Using
\er{SD} and \er{trn} we obtain \er{SD2}.
 $\BBox$

 {\bf Proof of Theorem \ref{T1}.}
If $q\in L^1(\R)\cap L^2(\R)$ and $q'\in L^1(\R)$ we get
$$
\begin{aligned}
\Tr Y_0^2=&\Tr VR_0VR_0=\Tr qS_0\overline{q}T_0 +\Tr \overline{q}T_0q S_0=2\Tr qS_0\overline{q}T_0=\\
=&-2\int_{0}^\infty q(x)e^{-2i\l x}\int_{x}^\infty e^{2i\l y}\overline{q}(y)dydx=\frac{1}{i\l}\int_0^\infty |q(x)|^2dx-\\
&\frac{1}{i\l}\int_0^\infty q'(x)e^{-2i\l x}\int_x^\infty e^{2i\l y}\overline{q}(y)dydx,
\end{aligned}
$$
which together with (\ref{SD2}) shows
$$
-\log D(\l)=\frac{1}{2i\l}\int_0^\infty|q(x)|^2dx + o(\l^{-1}) \qqq \as \qq |\l|\to \iy,
\qq \l\in \ol\C_+.
$$
This implies (\ref{2}) in Theorem \ref{T1} if we show Formula (\ref{a=D}) in Theorem \ref{T1}.
We prove the following: Let $q\in L^1(\R)\cap L^2(\R).$ Then we have \\ i)
${\displaystyle D\in AC(\C_+,\C),\qq\det \cS(\l)=\frac{D(\l-i 0)}{D(\l+i 0)},\qq\forall\l\in\R.}$\\
ii)
$
D=a.
$
\\

 i) We use arguments from \cite{IK11}. Let $\l\in\C_+.$  Denote $\cJ_0(\l)=I+Y_0(\l),$
$\cJ(\l)=I-Y(\l).$ Then $\cJ_0(\l)\cJ(\l)=I$ due to (\ref{2.5}). Now, put
$S_0(\l)=\cJ_0(\overline{\l})\cJ(\l).$ Then we have
$$
S_0(\l)=I-\left(
Y_0(\l)-Y_0(\overline{\l})\right)\left(I-Y(\l)\right).
$$
Now, by the
Hilbert identity,
$$
Y_0(\l)-Y_0(\overline{\l})=(\l-\overline{\l})V_2R_0(\l)R_0(\overline{\l})V_1
$$
 is trace class, and by taking the limit $\l\pm i\epsilon,$ $\l\in\R,$ $\epsilon\rightarrow 0,$ we get
$$\det S_0(\l)=\det \cS(\l),\qq \l\in \R.$$

Let $z=i\tau,$ $\tau\in\R_+$ and $\cD=\det (\cJ_0(\l)\cJ(z)),$
$\l\in\C_+.$ \\
 It is well defined as $\cJ_0(.)\cJ(z)-I\in
AC(\C_+;\cB_1).$ The function $\cD(\l)$ is entire in $\C_+$ and
$\cD(z)=I.$ We put $$f(\l)=\frac{D(\l)}{D(z)}e^{\Tr
(Y_0(\l)-Y_0(z))},\qq \l\in\overline{\C_+},$$ where
$$D(\l)=\det\left[ (I+Y_0(\l)) e^{-Y_0(\l)}\right].$$ We have
  $\cD(\l)=f(\l),$ $\l\in \C^+.$
Now, using that $\cJ_0(\l)\cJ(\l)=I,$ we get
$$
\det S_0(\l)=\det J_0(\overline{\l})J(z)\cdot
\det(\cJ(z)^{-1}\cJ(\l)=\frac{\cD(\overline{\l})}{\cD(\l)}=
\frac{D(\overline{\l})}{D(\l)}e^{\Tr (Y_0(\overline{\l})-Y_0(\l))}.
$$
As by (\ref{l-trace0}) we have $\Tr(Y_0(\l+i0)-Y_0(\l-i0))=0$ for
$\l\in\R,$   then we get
$$\det \cS(\l)=\lim_{\epsilon\downarrow 0}
\frac{D(\l-i\epsilon)}{D(\l+i\epsilon)},\qq\l\in\R.$$

ii) Now, we obtained
 \[
 \lb{Da}
 {\ol D(\l+i0)\/D(\l+i0)}={\ol a(\l+i0)\/a(\l+i0)}\qqq \qqq \forall
 \ \l\in \R.
 \]
 Moreover, due to \er{ED10} and \er{estimate} we have also
\[
\begin{aligned}
 \lb{Da1}
&D(\l)\to 1,\qqq\qqq a(\l)\to 1 \qqq {\rm as}\ \  \Im \l\to \iy,\\
&D(\l)\ne 0,\qqq\qqq a(\l)\ne 0 \qqq \forall
 \ \l\in \ol\C_+.
\end{aligned}
\]
Thus we can define uniquely the functions
$
\log D(\l),$  $\log a(\l)$ $\forall\,\l\in \ol\C_+,
$
by the conditions
\[
\lb{Da2} \log D(\l)\to 0,\qqq \log a(\l)\to 0 \qqq {\rm as}\ \  \Im
\l\to \iy.
\]
This and \er{Da1} imply
\[
\lb{Da3} e^{-2i\arg D(\l+i0)}=e^{-2i\arg a(\l+i0)},\qq \forall \
\l\in\R.
\]
 The functions $\log D(\l), \log a(\l)$ are
analytic in $\C_+$ and continuous up to the real line. Then $\arg
D(\l+i0)=\arg a(\l+i0)+2\pi N$ for all $\l\in\R$ and for some
integer $N\in \Z$. We define a new function $F(\l)=\arg D(\l)-\arg
a(\l)$ for all $\l\in\ol\C_+$. This function satisfies
$$
F(\l+i0)=2\pi N,\qqq \forall\ \l\in\R,\qqq {\rm and}\qqq F(\l)\to 0
\qqq {\rm as}\ \ \Im \l\to \iy.
$$
Thus $F\equiv 0$ and $D\equiv a.$

Now, we prove (\ref{Hardy}). As by ii), we have $a=D,$ it is enough to consider $\log a(\cdot).$ Note that $|\log(a(\l))| \leq |a(\l)-1|.$
Recall that  we have $|a(\l)-1|\leq C$ uniformly in   $\l\in\C_+$ (see (\ref{estimate})) and from (\ref{a_in_L2}) it follows that $\int_\R|a(x)-1|^2dx\equiv M <\infty.$ Now, the Plancherel-P{\'o}lya theorem (see \cite{L93} )
yields
$\int_\R|a(x+iy)|^2 dx \leq M <\infty$ uniformly in $y>0.$

\hfill $\BBox$

{\bf Proof of Theorem \ref{T4}.} We suppose that $q$ satisfies Condition A.
Recall that from Corollary \ref{Ttrace}, (\ref{tr2}), it follows that $R(\l)-R_0(\l)$ is trace class.
Therefore
 $f(H)-f(H_0)$ is trace class for any $f\in \mS,$ where $\mS$ is the Schwartz class of all rapidly decreasing functions, and
the Krein's trace formula is valid (general result):
$$
\Tr (f(H)-f(H_0))=\int_\R\xi(\l)f'(\l)d\l,\qq f\in \mS,$$ where
$\xi(\l)=\frac{1}{\pi}\phi_{\rm sc}(\l)$ is the spectral shift function and $\phi_{\rm sc}(\l)=\arg a(\l)=\frac{i}{2}\log\det \cS$  is the scattering phase.

As for $\l\in\R,$
$$
\det \cS=\frac{\overline a}{a},$$ then we have also
$$\frac{(\det \cS)'}{\det \cS}= -2i\Im\frac{a'(\l)}{a(\l)}.
$$

By using the Hadamard factorization (\ref{Had-fact}) from Proposition  \ref{Prop2} we get
\[\lb{had}
\frac{a'(\l)}{a(\l)}=i\g+\lim_{r\to +\infty} \sum_{|\l_n|\le
r}\frac{1}{\l-\l_n},\]

Now, using (\ref{had}) , we get
$$\Tr (f(H)-f(H_0))=\frac{1}{2\pi i}\int_\R f(\l)\frac{(\det \cS)'}{\det \cS}d\l=-\frac{1}{\pi}\lim_{r\to +\infty} \sum_{|\l_n|\le
r}\int_\R f(\l)\Im \frac{1}{\l-\l_n}d\l$$ and
$$\Tr (f(H)-f(H_0))=-\frac{1}{\pi}\lim_{r\to +\infty} \sum_{|\l_n|\le
r}\int_\R f(\l)\frac{\Im\l_n}{|\l-\l_n|^2}d\l,$$ recovering the Breit-Wigner profile ${\displaystyle -\frac{1}{\pi}\frac{\Im\l_n}{|\l-\l_n|^2}.}$ The sum is converging absolutely by (\ref{sumcond}).

Now using (\ref{ED1}), (\ref{2.10}), (\ref{2.5}) and
$R(\l)-R_0(\l)=-R_0V_1(I+Y_0(\l))^{-1}V_2R_0(\l)$
 we get $\frac{d}{d\l} \log D(\l)=$
\[\lb{DD}\frac{D'(\l)}{D(\l)}=-\Tr Y(\l)Y_0'(\l)=-\Tr [Y_0'(\l)-(I-Y(\l))Y_0'(\l)]=-\Tr (R(\l)-R_0(\l)).\]
Recall that if potential $q$ satisfies Condition A then $D=a\in \cE_1(2\gamma)$ (Proposition \ref{Prop2})
and \[\lb{D1D} \frac{D'(\l)}{D(\l)}=\frac{a'(\l)}{a(\l)}.\]

Now, using (\ref{D1D}) and the Hadamard factorization (\ref{had}) in (\ref{DD})
we get the trace formula
$$ \Tr (R(\l)-R_0(\l))=-i\g-\lim_{r\to +\infty} \sum_{|\l_n|\le
r}\frac{1}{\l-\l_n}$$ with uniform convergence in every disc or bounded subset of the plane.

This proves (\ref{TrF1}) and (\ref{sc_phase}) in Theorem \ref{T4}.\hfill
\BBox

\no  {\bf Acknowledgments.} \small
Authors are grateful to the institute Mittag-Leffler, Djursholm, where this paper was completed.
Various parts of this paper were written during Evgeny Korotyaev's stay
at the faculty of Technology and Society, Malm{\"o} University   and he acknowledges the faculty for its hospitality and the  Wenner--Gren foundation for the funding of his stay.
Evgeny Korotyaev's study was supported by the Ministry of education and science of Russian Federation, project   07.09.2012  No  8501 No «2012-1.5-12-000-1003-016»
and the RFFI grant "Spectral and asymptotic methods
for studying of the differential operators" No 11-01-00458.


\end{document}